\documentclass[prl,aps,twocolumn,showpacs]{revtex4}
\usepackage{epsfig}
\include{psfig}

\begin{document}

\title {Orbital entanglement and violation of Bell inequalities in
mesoscopic conductors}

\author{P. Samuelsson, E.V. Sukhorukov and M. B\"uttiker}
\affiliation{D\'epartement de Physique Th\'eorique, Universit\'e de
Gen\`eve, CH-1211 Gen\`eve 4, Switzerland.}  
\pacs{74.50.+r, 73.50.Td, 73.23.Ad}
\begin{abstract}
We propose a spin-independent scheme to generate and detect
two-particle entanglement in a mesoscopic normal-superconductor
system. A superconductor, weakly coupled to the normal conductor,
generates an orbitally entangled state by injecting pairs of electrons
into different leads of the normal conductor. The entanglement is
detected via violation of a Bell inequality, formulated in terms of
zero-frequency current cross-correlators. It is shown that the Bell
inequality can be violated for arbitrary strong dephasing in the
normal conductor.
\end{abstract}
\maketitle Entanglement is one of the most intriguing features
predicted by quantum theory \cite{EPR}. It leads to correlation
between distant particles, which can not be described by any local,
realistic theory \cite{Bell}. This nonlocal property of entanglement
has been demonstrated convincingly in optics \cite{Gisin1}, where
entangled pairs of photons have been studied over several
decades. Apart from the fundamental aspects, there is a growing
interest in using the properties of entangled particles for applied
purposes, such as quantum cryptography \cite{GisinRMP} and quantum
computation \cite{SteaneRPP}.

Recently, much interest has been shown for entanglement of electrons
in solid state devices. A controlled generation and manipulation of
electronic entanglement is of importance for a large scale
implementation of quantum information and computation schemes.
Electrons are however, in contrast to photons, massive and
electrically charged particles, which raises new fundamental questions
and new experimental challenges. Existing suggestions are based on
creating \cite{Recher,Bena}, manipulating and detecting
\cite{Detect,Les,Cht01} spin-entangled pairs of electrons. This
requires experimental control of individual spins via spin filters or
locally directed magnetic fields on a mesoscopic scale. Here we
propose a spin-independent scheme for creating and detecting orbital
entanglement in a mesoscopic normal-superconductor system.

\begin{figure}[h]
\centerline{\psfig{figure=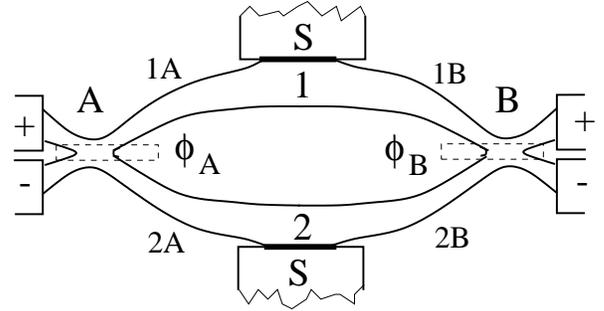,width=0.9\linewidth}}
\caption{The system: A single superconductor
(S) is connected to four normal arms via two tunnel barriers $1$ and
$2$ (thick black lines). The arms are joined pairwise in beam splitters
A and B and end in normal reservoirs $+$ and $-$.}
\label{junction}
\end{figure}

We show that a superconductor, weakly coupled to a normal conductor
(see Fig.\ \ref{junction}), creates an orbitally entangled state by
emitting a coherent superposition of pairs of electrons into different
leads of the normal conductor. The zero-frequency correlation between
currents flowing into different normal reservoirs is shown to be
equivalent to a pair coincidence measurement: only correlations
between the electrons from the same entangled pair contribute. As a
consequence, a standard Bell Inequality (BI) can be directly
formulated in terms of the zero-frequency current correlators. We find
that a violation of the BI, demonstrating the entanglement of the pair
state, can be obtained for arbitrary dephasing in the normal
conductor.

We first consider a simplified version of the system (see
Fig. \ref{junction}), a more detailed discussion is given below. A
single \cite{supcom} superconductor (S) is weakly coupled to a normal
conductor, a ballistic two-dimensional electron gas, via two tunnel
barriers $1$ and $2$ with transparency $\Gamma \ll 1$. The normal
conductor consists of four arms, $1A,1B,2A$ and $2B$, with equal
lengths $L$. The arms $1A$ and $2A$ ($1B$ and $2B$) are crossed in a
controllable beam splitter $A(B)$, parameterized via the angle
$\phi_A$($\phi_B$), and then connected to normal reservoirs $+A$ and
$-A$ ($+B$ and $-B$).
The beam splitters \cite{Oliver2} are assumed to support only one
propagating mode. The states $|+,\eta \rangle$ and $|-,\eta \rangle$
for electrons going out into the normal reservoirs and the states
$|1,\eta\rangle$ and $|2,\eta \rangle$ of the electrons emitted from
superconductor are related via a scattering matrix:
\begin{eqnarray}
\left[\begin{array}{c} |+,\eta\rangle \\ |-,\eta \rangle \end{array}\right]=
\left(\begin{array}{cc} \cos \phi_{\eta} & -\sin
\phi_{\eta} \\ \sin \phi_{\eta} & \cos
\phi_{\eta} \end{array}\right) \left[\begin{array}{c}
|2,\eta \rangle \\ |1,\eta \rangle \end{array}\right], \eta=A,B.
\label{smat}
\end{eqnarray}
The angles $\phi_A$ and $\phi_B$ can be tuned between $0$ and $\pi/2$
by tuning the beam splitter from ``open'', when the electrons are
transmitted through from $1(2)$ to $-(+)$, to ``closed'', when the
electrons are completely reflected, from $1(2)$ to $+(-)$. We consider
the low temperature limit, $kT \ll eV$. A negative voltage $-eV$ is
applied to all the normal reservoirs and the superconductor is
grounded. The voltage $eV$ is smaller than the superconducting gap
$\Delta$, so no single particle transport takes place. It is further
assumed that the size of the system is smaller than the phase breaking
length.

We first present a simple and transparent explanation of how the
entanglement is generated and detected, a rigorous derivation follows
below. The superconductor emits pairs of particles into the normal
arms. Since the superconductor is a single object, coherent on a
macroscopic scale, the state of an emitted pair is a linear
superposition of states corresponding to a pair emitted through
barrier $1$ and $2$. The emitted pair can either split, with one
electron going to each beam splitter $A$ and $B$, or both electrons
can go to the same beam splitter. However, the latter process does not
contribute \cite{Samuelsson} to the quantity of interest, the current
cross-correlations, to leading order in pair emission (Andreev
reflection) probability, proportional to $\Gamma^2$. The relevant part
of the state of the emitted pair can thus be written as
\begin{equation}
|\Psi\rangle=|\Psi_{12}\rangle \otimes|\Psi_{AB}\rangle, \hspace{5mm}|\Psi_{12}\rangle=\left(|11\rangle+|22\rangle\right)/\sqrt{2},
\label{state}
\end{equation}
a product of a state $|\Psi_{12} \rangle$, orbitally entangled with
respect to emission across barrier $1$ and $2$ (called $12$-space
below), and a state $|\Psi_{AB}\rangle$, describing one electron going
towards $A$ and one towards $B$, containing all additional
information, such as energy and spin dependence. The beam splitters
$A$ and $B$, parameterized via the angles $\phi_A$ and $\phi_B$, rotate
the state $|\Psi_{12}\rangle$ [see Eq. (\ref{smat})].

The entanglement is detected by violation of a BI. We recall that in
the original formulation \cite{Bell}, a source emitting spin-$1/2$
singlets was considered. The BI, as formulated in Ref. \cite{CHSH},
\begin{eqnarray}
S&\equiv&|E(\phi_A,\phi_B)-E(\phi_A,\phi_{B}')\nonumber \\ 
&+&E(\phi_{A}',\phi_B)+E(\phi_{A}',\phi_{B}')|\leq 2,
\label{BI1}
\end{eqnarray}
is expressed in terms of spin correlation functions \cite{Aspect}
\begin{equation}
E(\phi_A,\phi_B)=P_{++}-P_{+-}-P_{-+}+P_{--}.
\label{BI2}
\end{equation}
Here $P_{\alpha\beta}(\phi_A,\phi_B)$ are the joint probabilities to
observe one particle in detector $A$ with a spin $\alpha=\pm$
[$+$($-$) denoting up (down)] along the $\phi_A$-direction, and the
other in detector $B$ with a spin $\beta=\pm$ along the
$\phi_B$-direction. The joint probabilities are given by
\begin{equation}
P_{\alpha\beta}(\phi_A,\phi_B)=1+\alpha\beta\cos[2(\phi_A-\phi_B)].
\label{corrsimp}
\end{equation}
Inserting the values for $P_{\alpha\beta}$ from Eq. (\ref{corrsimp})
into Eq. (\ref{BI2}), we get
$E(\phi_A,\phi_B)=\cos[2(\phi_A-\phi_B)]$. We then directly see that
for e.g. angles $\phi_A=\pi/8,\phi_B=\pi/4,\phi_A'=3\pi/8$ and
$\phi_B'=\pi/2$, the BI in Eq. (\ref{BI1}) is maximally violated,
i.e. we get $S=2\sqrt{2}$.

In our orbital setup (see Fig \ref{junction}), it is clear from
Eq. (\ref{state}) that the $12$-space plays the role of a pseudo-spin
space and the normal reservoirs act as detectors. We can thus, in
direct analog to Ref.\ \cite{Bell}, formulate a BI in terms of any
observable which is directly proportional to the corresponding joint
probability $P_{\alpha\beta}(\phi_A,\phi_B)$ for our state
$|\Psi_{12}\rangle$ [here $\alpha,\beta=\pm$ denote the reservoirs,
see Fig. \ref{junction}]. We find below that the zero-frequency
current cross-correlator is given by
\begin{eqnarray}
S_{\alpha\beta}\equiv 2\int_{-\infty}^{\infty}dt\langle \delta \hat
I_{\alpha A}(t) \delta \hat I_{\beta
B}(0)\rangle=P_0P_{\alpha\beta}(\phi_A,\phi_B),
\label{pendep}
\end{eqnarray}
i.e. directly proportional to the joint probability distribution.
Here $P_0=e^3V\Gamma^2/2h$, and $\delta \hat I_{\alpha\eta}(t)=\hat
I_{\alpha\eta}(t)-\langle \hat I_{\alpha\eta}\rangle$ is the
fluctuating part of the current $\hat I_{\alpha\eta}(t)$ in reservoir
$\alpha\eta$. This leads to the important result that the Bell
inequality in Eq.\ (\ref{BI1}) can be directly formulated in terms of
the zero-frequency current correlators in Eq.\ (\ref{pendep}). We note
\cite{Aspect} that when inserting $S_{\alpha\beta}$ directly into the
correlation functions Eq.\ (\ref{BI2}) (i.e. substituting
$P_{\alpha\beta}$ with $S_{\alpha\beta}$), we must divide by the sum
of all correlators $S_{++}+S_{+-}+S_{-+}+S_{--}=P_0$, which just
eliminates $P_0$.

\begin{figure}[h]
\centerline{\psfig{figure=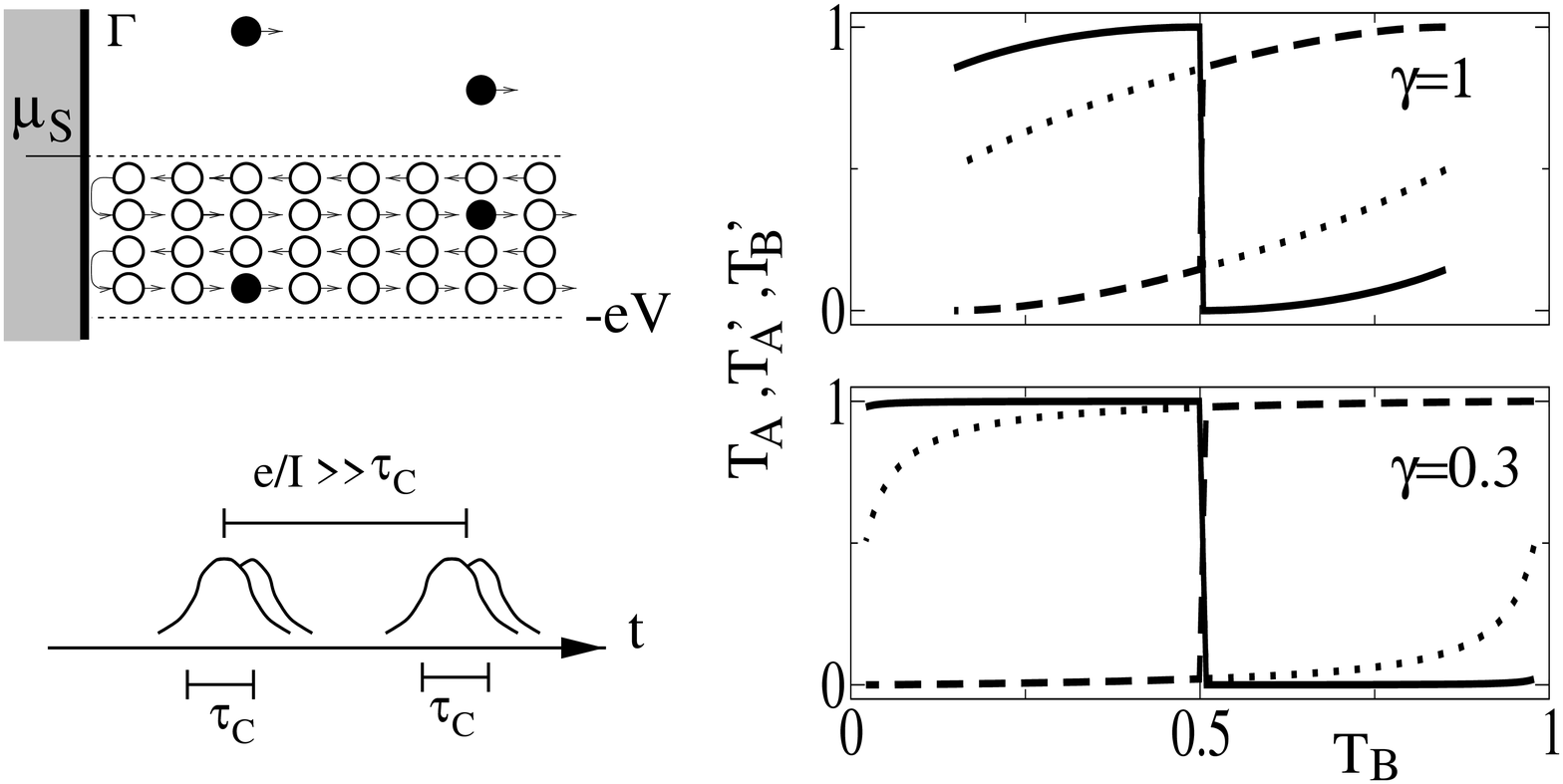,width=0.95\linewidth}}
\caption{Left, upper: In the filled stream of incoming holes (empty
circles) from the normal reservoir, occasionally a hole is
back-reflected as an electron (solid circles). The ``missing hole''
(i.e an electron) and the Andreev reflected electron constitute the
pair emitted from the superconductor. Left, lower: The correlation
time $\tau_c=\hbar/eV$ (width of the wave packet) and the average time
between emission of two subsequent pairs
$e/I=\hbar/eV\Gamma^2\gg\tau_c$ of the current correlator. The small
time difference $\hbar/\Delta \ll \tau_c$ between the emissions of the
two electrons in the pair is shown as a split of the
wavepacket. Right: The transmission probabilities $T_{A}$
(dashed),$T_A'$ (dotted) and $T_B'$ (solid) as a function of $T_{B}$
[$T_{\eta}=\cos^2(\phi_{\eta})]$, giving optimal violation of the Bell
inequalities for dephasing parameters $\gamma=1$ and $\gamma=0.3$.}
\label{andreev}
\end{figure}

The simple result in Eq.\ (\ref{pendep}) can be understood by
considering the properties of the time-dependent correlator $\langle
\delta \hat I_{\alpha A}(t) \delta \hat I_{\beta B}(0)\rangle$. It is
finite only for times $t\alt \tau_c$, where $\tau_c=\hbar/eV$ is the
correlation time of the emitted pair (see Fig.\ \ref{andreev}). In the
tunneling limit under consideration, $\Gamma \ll 1$, the correlation
time is much smaller than the average time between the arrival of two
pairs $e/I \sim \hbar/eV\Gamma^2$. As a result, only the two electrons
within a pair are correlated with each other, while electrons in
different pairs are completely uncorrelated. Thus, the zero frequency
current correlator in Eq.\ (\ref{pendep}) is just a coincidence
counting measurement running over a long time, collecting statistics
over a large number of pairs \cite{Chetcom}.

For a rigorous derivation of the above result, we first discuss the
role of the superconductor as an emitter of pairs of orbitally
entangled electrons. Since the system is phase coherent, we can work
within the scattering approach to normal-superconducting systems
\cite{Datta,tuncom}. The starting point is the many-body state of the
normal reservoirs, describing injection of hole quasiparticles at
energies from $0$ to $eV$ (suppressing spin notation),
\begin{equation}
|\Psi_{in}\rangle=\prod_{0<E<eV}\gamma^{\dagger}_{1A}(E)\gamma^{\dagger}_{2A}(E)\gamma^{\dagger}_{1B}(E)\gamma^{\dagger}_{2B}(E)|0\rangle,
\label{ehmbstate12}
\end{equation}
where the ground state $|0\rangle$ is the vacuum for quasiparticles in
the normal reservoir. The operator $\gamma_{1A}^{\dagger}(E)$ creates
a hole plane wave at energy $E$ (counted from the superconducting
chemical potential $\mu_S$) in lead $1A$, going out from the normal
reservoirs towards the superconductor, and similarly for the other
operators. The commutation relations for the $\gamma$-operators are
standard fermionic.

To obtain the state of the quasiparticles going out from the
superconductor, $|\Psi_{out}\rangle$, we note that the operators
creating and destroying quasiparticles going out from the
superconductor (towards the normal reservoirs) are related
\cite{Datta,Buttiker} to the operators of the incoming quasiparticles
via a scattering matrix. The scattering amplitude for a hole injected
in arm $1A$, to be back-reflected as an electron in arm $2A$, is
denoted $r_{2A,1A}^{eh}$ and similarly for the other scattering
amplitudes. In the tunnel limit under consideration, the amplitude to
backscatter as the same type of quasiparticle is
$r^{hh}_{j\eta,j'\eta'} \sim 1$. The amplitude for Andreev reflection,
$r^{eh}_{j\eta,j'\eta'}$, is given by $-i \Gamma/4$ (independent on
energy). Processes where a hole incident on barrier $1(2)$ is
backscattered as an electron at barrier $2(1)$, i.e. when a pair in
the superconductor breaks up, are exponentially suppressed with the
distance between the two emission points\ \cite{Recher} and can be
neglected in the present setup.

The tunneling limit $\Gamma \ll 1$ makes it relevant to change the
perspective from a quasiparticle picture to an all-electron
picture. An Andreev reflection, occuring with a small probability
$\Gamma^2$, can be considered as a perturbation with respect to the
ground state $|\bar 0\rangle$ in the normal reservoirs (a filled Fermi
sea of electrons at energies $E<-eV$). It creates an excitation
consisting of a pair of electrons (see Fig.\ \ref{andreev}). Formally,
performing a Bogoliubov transformation
(i.e. $\gamma_{j\eta}(E)=c^{\dagger}_{j\eta}(-E)$, with
$c^{\dagger}_{j\eta}(E)$ being a standard electron creation
operator), the state of the quasiparticles going out from the
superconductor becomes \cite{Optcom} to first order in $\Gamma$
\begin{eqnarray}
|\Psi_{out}\rangle=|\bar 0\rangle+|\tilde \Psi\rangle+|\Psi\rangle,
\label{entstate}
\end{eqnarray}
where the states
\begin{eqnarray}
|\tilde \Psi\rangle&=&\frac{i\Gamma}{4}\int_{0}^{eV}dE
 \left[\sum_{j=1,2}\sum_{\eta=A,B}c_{j\eta}^{\dagger}(E)c_{j\eta}(-E)
 \right]|\bar 0\rangle, \nonumber \\
 |\Psi\rangle&=&\frac{i\Gamma}{4}\int_{-eV}^{eV}dE\left[c_{1A}^{\dagger}(E)c_{1B}^{\dagger}(-E)
 \right. \nonumber \\ && \hspace{1cm} +
 \left. c_{2A}^{\dagger}(E)c_{2B}^{\dagger}(-E)\right]|\bar 0\rangle,
\label{entNSwavefcn12}
\end{eqnarray}
describe orbitally entangled electron ``wave packet'' pairs, i.e. a
superposition of pairs at different energies \cite{sukhcom}. In a
first quantized notation, the state $|\Psi \rangle$ is just the state
in Eq.\ (\ref{state}). We emphasize that the change from a
quasiparticle to an all-electron picture, providing a clear picture of
the entanglement, does not alter the physics.

The detection of the entanglement is done via the zero frequency
current cross-correlators $S_{\alpha\beta}$ in Eq.\ (\ref{pendep}). To
obtain $S_{\alpha\beta}$, we insert the second quantization expression
for the current operator \cite{Buttiker}, $\hat
I_{\alpha\eta}(t)=(e/h)\int dEdE'\mbox{exp}[i(E-E')t/\hbar]
c_{\alpha\eta}^{\dagger}(E')c_{\alpha\eta}(E)$, into
Eq. (\ref{pendep}) and average with respect to the state
$|\Psi\rangle$ in Eqs.\ (\ref{entstate}) and
(\ref{entNSwavefcn12}). The beam splitters are taken into account by
relating the $c_{j\eta}$-operators in arms $1A$ and $2A$ ($1B$ and
$2B$) to the $c_{\alpha\eta}$-operators in the reservoirs $+A$ and
$-A$ ($+B$ and $-B$) via the scattering matrix in
Eq. (\ref{smat}). The average current, equal in all arms $\alpha\eta$,
is $\langle I_{\alpha\eta}\rangle\equiv I=(e^2/2h)\Gamma^2V$,
independent of the beam splitter transparency.

The ground state $|\bar 0\rangle$ in Eq. (\ref{entstate}) does not
contribute to the correlator. Moreover, calculating the current
correlator we find that the state $|\tilde \Psi\rangle$, describing
two electrons emitted into the same normal lead, only contributes
\cite{Samuelsson} to the cross correlator at order $\Gamma^4$ and can
thus be neglected. From Eq.\ (\ref{pendep}) we then find that
$S_{\alpha\beta}(\phi_A,\phi_B)=P_0P_{\alpha\beta}(\phi_A,\phi_B)$,
where $P_{\alpha\beta}$ is given in Eq. (\ref{corrsimp}), just as
announced. We note that it is the structure of $|\Psi\rangle$ in
$12$-space, i.e. $|\Psi_{12}\rangle$, that determines the angle
dependence of $P_{\alpha\beta}$, all properties of $|\Psi_{AB}\rangle$
just gives rise to the constant $P_0$. Moreover, the calculations show
that the time dependent correlator $\langle \delta \hat I_{1\alpha}(t)
\delta \hat I_{2\beta}(0)\rangle$ vanishes as $(\tau_c/t)^2$ for $t
\gg \tau_c$. As is pointed out above, $S_{\alpha\beta}(\phi_A,\phi_B)$
can thus be inserted in Eq. (\ref{BI2}) and be used to violate the BI
in Eq. (\ref{BI1}).

Until now, ideal conditions have been considered. One possible
source of disturbances is dephasing. In our system, dephasing can
quite generally be expressed in terms of a density matrix
$\rho=[|11\rangle\langle 11|+|22\rangle \langle
22|+\gamma(|11\rangle\langle 22|+|11\rangle \langle 22|)]/2$, where the
off-diagonal elements, giving rise to the entanglement, are suppressed
by a phenomenological dephasing parameter $0 \leq \gamma \leq 1$. The
correlators $E(\phi_A,\phi_B)$ in Eq. (\ref{BI2}) then take the form
\begin{equation}
E(\phi_A,\phi_B)=\cos (2\phi_A)\cos (2\phi_B)+\gamma\sin (2\phi_A)\sin
(2\phi_B).
\label{eimp}
\end{equation}
By adjusting the four angles $\phi_A,\phi_A',\phi_B$ and $\phi_B'$
we find the maximal Bell parameter in Eq. (\ref{BI1}) is
\begin{equation}
S=2\sqrt{1+\gamma^2}
\label{BI3}
\end{equation}
which violates the BI for any $\gamma>0$. The optimal violation
angles, all in the first quadrant, are
$\tan(2\phi_A)=-\gamma\cot(\phi_S)$,
$\tan(2\phi_{A}')=\gamma\tan(\phi_S)$ and
$\tan(\phi_B-\phi_{B}')=\mbox{sign}[\cos(2\phi_A)][(\tan^2(\phi_S)+\gamma^2)/(\gamma^2\tan^2(\phi_S)+1)]^{1/2}$,
where $\phi_S=\phi_B+\phi_{B}'$ can be chosen at will. The
corresponding transmission probabilities
$T_{\eta}=\cos^2(\phi_{\eta})$ are shown for $\gamma=1$ and
$\gamma=0.3$ in Fig.\ \ref{andreev}.

The BI can thus in principle be violated for any amount of
dephasing. However it might be difficult to produce beam splitters
which can reach all transmission probabilities between $0$ and
$1$. This is not a serious problem in the absence of dephasing,
$\gamma=1$, a violation can be obtained for a large, order of unity,
fraction of the ``transmission probability space''. However, in the
limit of strong dephasing, $\gamma \ll 1$, the set of transmission
probabilities for optimal violation contain transmissions close to
both $0$ and $1$, see Fig. \ref{andreev}. Expecting unity transmission
to be most complicated to reach experimentally, we note that by
instead choosing transmission probabilities $T_A=T_B=0$, $T_B'=1/2$
and $T_A' \ll \gamma$, the inequality in Eq. (\ref{BI1}) becomes
$2|1+\gamma T_A'|\leq 2$. This gives a violation, although not
maximal, for all $\gamma\ll 1$.

Apart from dephasing there are several other effects such as
additional scattering phases, impurity scattering or asymmetric tunnel
barriers, which might alter the possibility to violate the BI. It
turns out that all these effects can be taken into account by
replacing $\gamma \rightarrow \gamma' \cos(\phi_0)$ in Eq.\
(\ref{BI3}), with the important conclusion that none of these effects
will destroy the possibility to violate the BI.

The phase factor $\phi_0$ is the sum of possible scattering phases
from the beam splitters [the scattering amplitudes in Eq.\
(\ref{smat}) are taken real], phases $\sim k_F \Delta L$ due to a
difference in length, $\Delta L$, between the normal arms (see
Fig. \ref{junction}), scattering phases from weak impurities and a
possible phase difference between the superconductor at the two tunnel
contacts $1$ and $2$. As a consequence e.g. the superconducting phase
(in a loop geometry) can be modulated to compensate for the other
phases.

The factor $\gamma'$ plays the same role as dephasing. One possible
contribution to $\gamma'$ is energy dependent phases which oscillate
rapidly on a scale of $eV$, suppressing the entangled part of the
current correlator. For different lengths of the normal arms, there is
always a phase $\sim E \Delta L/\hbar v_F$. This phase can however be
neglected for $\Delta L \ll \hbar v_F/eV$, which for $eV \ll \Delta$
is fulfilled for $\Delta L$ smaller than the superconducting
coherence length $\hbar v_F/\Delta$. Another possibility is that, due
to asymmetries of the tunnel barriers $\Gamma_1 \neq \Gamma_2$, the
amplitude for the process where the pair is emitted to $|11\rangle$ is
different from the process where it is emitted to $|22\rangle$. This
gives rise to a state, in 12-space,
($\Gamma_1|11\rangle+\Gamma_2|22\rangle)/\sqrt{\Gamma_1^2+\Gamma_2^2}$).
In this case \cite{Gisin91} $\gamma'=
2\Gamma_1\Gamma_2/(\Gamma_1^2+\Gamma_2^2)$. Thus, it is in
principle possible to violate BI for arbitrary asymmetry.

We finally point out that the constraint on single mode beam splitter
can not easily be relaxed. A multi-mode beam splitter will, in the
suggested setup, probably have a different transparency dependence for
the different modes. This decreases the space of accessible angles and
will eventually make the violation of the BI impossible.

In conclusion, we have investigated a spin-independent scheme to
generate and detect two-particle orbital entanglement in a mesoscopic
normal-superconductor system. The cross-correlator between the
currents in the two leads depends in a nonlocal way on transparencies
of beam splitters in the two leads. These nonlocal correlations arises
due to the entanglement of the injected pair. For appropriate choices
of transparencies, the correlators give a violation of a BI for
arbitrary strong dephasing.

We acknowledge discussions with Fabio Taddei and Valerio Scarani. This
work was supported by the Swiss National Science Foundation and the
program for Materials with Novel Electronic Properties.


\begin{thebibliography}{999}
\bibitem{supcom} 
The two parts of the superconductor can be contacted e.g. via a loop.
\bibitem{Chetcom} 
In contrast to Ref. \cite{Cht01} there is no
constraint on the measurement time in our scheme.
\bibitem{tuncom} The same result can be obtained using the tunneling
Hamiltonian formalism in ref. \cite{Recher}.
\bibitem{sukhcom} 
In contrast to Refs. \cite{Detect}, in our setup there is no
additional constraint on the energy spectrum of the entangled electrons.
\end{thebibliography}
\end{document}